\documentstyle[12pt,epsf]{article}
\begin{document}
\bibliographystyle{unsrt}
\noindent
\begin{center}
{\bf The Mode-Coupling Theory of the Glass Transition}\\
Walter Kob\footnote{
Electronic mail: kob@moses.physik.uni-mainz.de\\
http://www.cond-mat.physik.uni-mainz.de/\~{ }kob/home\_kob.html\\
To appear in {\it Experimental and Theoretical Approaches to
Supercooled Liquids: Advances and Novel Applications} Eds.: J. Fourkas,
D. Kivelson, U. Mohanty, and K. Nelson (ACS Books, Washington, 1997)
}\\
Institute of Physics, Johannes Gutenberg-University, Staudinger Weg 7,\\
D-55099 Mainz, Germany
\end{center}

\vspace*{7mm}
\par
\noindent
\begin{center}
\begin{minipage}[h]{122mm}
We give a brief introduction to the mode-coupling theory of the glass
transition, a theory which was proposed a while ago to describe the
dynamics of supercooled liquids. After presenting the basic equations
of the theory, we review some of its predictions and compare these with
results of experiments and computer simulations. We conclude that the
theory is able to describe the dynamics of supercooled liquids in
remarkably great detail.
\end{minipage}
\end{center}

\vspace*{5mm}
\par
\noindent
The dynamics of supercooled liquids and the related phenomenon of the
glass transition has been the focus of interest for a long
time~\cite{books_on_glasses}. The reason for this is the fact, that if
a glass former is cooled from its melting temperature to its glass
transition temperature $T_g$, it shows an increase of its relaxation
time by some 14 decades without a significant change in its structural
properties, which in turn poses a formidable and exciting challenge to
investigate such a system experimentally as well as theoretically.
Despite the large efforts that have been undertaken to study
this dramatic growth in relaxation time, even today there is still
intense dispute on what the underlying mechanisms for this increase
actually is. Over the course of time many different theories have been
put forward, such as, to name a few popular ones, the entropy theory by
Adams, Gibbs and Di Marzio~\cite{adam_gibbs}, the coupling-model
proposed by Ngai~\cite{ngai92}, or the mode-coupling theory (MCT) by
G\"otze and Sj\"ogren~\cite{mct_rev}. The approach by which these
theories explain the slowing down of the supercooled liquid with
decreasing temperature differs radically from case to case. In the
entropy theory it is assumed, e.g., that the slowing down can be
understood essentially from the thermodynamics of the system, whereas
MCT puts forward the idea that at low temperatures the nonlinear
feedback mechanisms in the microscopic dynamics of the particles become
so strong that they lead to the structural arrest of the system.

One of the most outstanding advantages of MCT over the other theories of
the glass transition is the fact that it offers a wealth of
predictions, some of which are discussed below, that can be tested
in experiments or computer simulations. This, and its
noticeable success, are probably the main reason why this theory has
attracted so much attention in the last ten years.  This is in contrast
to most other theories which make far fewer predictions and for which
it is therefore much harder to be put on a solid experimental
foundation.  This abundance of theoretical predictions has, of course,
its price, in that MCT is a relatively complex theory.  Therefore it is
not surprising that doing {\it quantitative} calculations within the
framework of MCT is quite complicated, although it is remarkable that
the theory gives well defined prescriptions how such calculations have
to be carried out and for simple models such computations have actually
been done.

The goal of this article is to give a brief introduction to the
physical background of MCT, then to review some of the main predictions
of the theory and to illustrate these by means of results from
experiments and computer simulations. In the final section we will
discuss a few recent developments of the theory and offer our view on
what aspect of the dynamics of supercooled liquids can be understood
with the help of MCT.

\section*{Mode-Coupling Theory: Background and Basic Equations}
\label{sec2}

In this section we give some historical background of the work that led
to the so-called mode-coupling equations, the starting point of MCT.
Then we present these equations and some important special
cases of them, the so-called schematic models.

In the seventies a considerable theoretical effort was undertaken in
order to find a correct quantitative description of the dynamics of
dense simple liquids. By using mode-coupling
approximations~\cite{kawasaki66}, equations of motion for density
correlation functions, described in more detail below, were
derived and it was shown that their solutions give at least a
semi-quantitative description of the dynamics of simple liquids in the
vicinity of the triple point. In particular it was shown that these
equations give a qualitatively correct description of the
so-called cage effect, i.e. the phenomenon that in a dense liquid a
tagged particle is temporarily trapped by its neighbors and that it
takes the particle some time to escape this cage.
 For more details the reader is referred to
Refs.~\cite{sjogren80_hansen_mcdonald86} and references therein. 

A few years later Bengtzelius, G\"otze and Sj\"olander (BGS) simplified
these equations by neglecting some terms which they argued were
irrelevant at low temperatures~\cite{bgs84}. They showed that the time
dependence of the solution of these simplified equations changes
discontinuously if the temperature falls below a critical value $T_c$.
Since this discontinuity was accompanied by a diverging relaxation time
of the time correlation functions, this singularity was tentatively
identified with the glass transition.

Let us be more specific: The dynamics of liquids is usually described
by means of $F(\vec{q},t)$, the density autocorrelation function for
wave vector $\vec{q}$, which is defined as:
\begin{equation}
F(\vec{q},t)=\frac{1}{N}\langle \delta\rho^{*}(\vec{q},t) 
\delta \rho(\vec{q},0)
\rangle \qquad \mbox{with} \qquad \rho(\vec{q},t)=\sum_{j=1}^{N}
\exp(i\vec{q}\cdot\vec{r}_j(t)),
\label{eq1}
\end{equation}
where $N$ is the number of particles and $\vec{r}_j(t)$ is the position
of particle $j$ at time $t$. The function $F(\vec{q},t)$, which is also
called intermediate scattering function, can be measured in scattering
experiments or in computer simulations and is therefore of practical
relevance.

For an isotropic system the equations of motion for $F(\vec{q},t)$ can
be written as
\begin{equation}
\ddot{F}(q,t)+\Omega^2(q)F(q,t)+\int_0^t\left[M^{0}(q,t-t')+ \Omega^2(q)
m(q,t-t')\right]\dot F(q,t')dt' = 0\quad .
\label{eq2}
\end{equation}
Here $\Omega(q)$ is a microscopic frequency, which can be computed from
the structure factor $S(q)$ via $\Omega^2(q)=q^2 k_B T/(mS(q))$ ($m$ is
the mass of the particles and $k_B$ Boltzmann's constant), the kernel
$M^{0}(q,t)$ describes the dynamics at short times and gives the only
relevant contribution to the integral at temperatures in the vicinity
of the triple point, whereas the kernel $m(q,t)$ becomes important at
temperatures where the system is strongly supercooled.  If we assume
that $M^{0}(q,t)$ is sharply peaked at $t=0$, and thus can be
approximated by a $\delta$-function, $M^{0}(q,t)=\nu(q) \delta(t)$, we
recognize from Eq.~(\ref{eq2}) that the equation of motion for $F(q,t)$
is the same as the one of a damped harmonic oscillator, but with the
additional complication of a retarded friction
which is proportional to $m(q,t)$.

It has to be emphasized that these equations of motion are {\it exact},
since the kernels $M^{0}(q,t)$ and $m(q,t)$ have not been specified
yet. In the approximations of the {\it idealized} version of
mode-coupling theory, the kernel $m(q,t)$ is expressed as a quadratic
form of the correlation functions $F(q,t)$, i.e.  $m(q,t)=
\sum_{\vec{k}+\vec{p}= \vec{q}} V(\vec{q};\vec{k},\vec{p})
F(k,t)F(p,t)$, where the vertices $V(\vec{q};\vec{k},\vec{p})$ can be
computed from $S(q)$.  With this approximation one therefore arrives at
a closed set of coupled equations for $F(q,t)$, the so-called
mode-coupling equations, whose solutions thus give the full time
dependence of the intermediate scattering function. These are the above
mentioned equations that were proposed and studied by BGS~\cite{bgs84}.
It is believed that they give a correct (self-consistent) description
of the dynamics of a particle at short times, i.e. when it is still in
the cage that is formed by its neighbors at time zero, and of the
breaking up of this cage at long times, i.e. up to the time scales when
the particle finally shows a diffusive behavior.

We also mention that in Eq.~(\ref{eq2}) the quantities $\Omega^2(q)$,
$M^0(q,t)$ and $V(\vec{q};\vec{k},\vec{p})$ depend on temperature. 
This temperature
dependence is assumed to be smooth throughout the whole temperature
range, an assumption which is supported by experiments and computer
simulations. Thus any singular behavior in the solution of the
equations of motion are due to their nonlinearity and not due to a
singularity in the input parameters. Since the strength of this
nonlinearity is related to the (temperature dependent) structure
factor, we thus see that the relevant temperature dependence of the
solution of Eq.~(\ref{eq2}) comes from the one of $S(q)$.

Due to the complexity of the mode-coupling equations their solutions can,
unfortunately, be obtained only numerically. Therefore BGS made the
approximation, which was proposed independently also by
Leutheusser~\cite{leutheusser84}, that the structure factor is given by
a $\delta$-function at the wave vector $q_0$, the location of the main
peak in $S(q)$~\cite{bgs84}. With this approximation Eq.~(\ref{eq2}) is
transformed into a single equation for the correlation function for
$q_0$, all the other equations vanish identically. By writing
$\Phi(t)=F(q_0,t)/S(q_0)$ we thus obtain
\begin{equation}
\ddot{\Phi}(t)+\Omega^2 \Phi(t) +\nu \dot{\Phi}(t)+\Omega^2 \int_0^t 
m[\Phi(t-t')]\dot{\Phi}(t') dt' =0 ,
\label{eq3}
\end{equation}
where $m[\Phi]$ is a low order polynomial in $\Phi$.  Such an equation
for a single (or at most very few) correlation function is called a
{\it schematic model}. Originally it was believed that such schematic
models reproduce some essential, i.e. universal, features of the full
theory. In the meantime it is understood what these universal features
of the full theory are. Therefore we know now, how to construct
schematic models, i.e. memory kernels $m[\Phi]$, so that they reproduce
the desired features of the full MCT equations.

We mentioned above that in the full mode-coupling equations the
relevant temperature dependence of the equations is given by the one of
the structure factor, which enters the coefficients of the quadratic
form of the memory kernel $m(q,t)$. In analogy to this one therefore
assumes that in the schematic models the coefficients of the polynomial
$m[\Phi]$ are also temperature dependent.

Since in these simplified models the details of all the
microscopic information has been eliminated they cannot be used to
understand experimental data {\it quantitatively}. However, their
greatly reduced complexity, as compared to the full equations, make
them amenable to analytic investigations from which many qualitative
properties of their solutions can be obtained.  Thus many of the
predictions of MCT stem from studying the solutions of the schematic
models, work that was done in the last ten years mainly by G\"otze,
Sj\"ogren and coworkers.  We will discuss these predictions in more
detail in the next section. For the moment we just mention briefly that
the analysis of the schematic models shows that if the nonlinearity,
given by the memory kernel $m[\Phi(t)]$, exceeds a certain threshold,
the solution of the equation does not decay to zero even at infinite
times.  This means that a density fluctuation that was present at time
zero does not disappear even at long times, i.e. the system is no
longer ergodic. Thus in this (ideal) case this dynamic transition can
be identified with the glass transition.

We mentioned earlier, that when BGS wrote down for the first time what
today are called the mode-coupling equations [Eq.~(\ref{eq2}), with
$m(q,t)$ given as a quadratic function of $F(q,t)$], they neglected in
these equations certain terms.  Later it was found that at very low
temperatures these terms do become important, since they lead to a
qualitatively different behavior of the time dependence of the solution
of the equations of motion~\cite{hopping}. In particular it was shown
that the above mentioned singularity in this solution disappears, i.e.
that even at low temperatures all the correlation functions decay to
zero at long times and that thus the system is always ergodic. Since
one of the mechanisms that can lead to the relaxation of the system,
and that is not taken into account by the idealized mode-coupling
equations, is a process in which one particle overcomes the walls of
its cage in an activated way, these processes are usually called {\it
hopping processes}. The version of MCT in which the effects of such
hopping processes are taken into account is called the {\it extended
version} of MCT. So far, however, the investigations of such processes
has been restricted to discuss solutions of mode-coupling equations in
which the hopping processes have been taken into account only in a
crude way, since even in these relatively simple cases the addition of
one additional parameter, the strength of the hopping processes, makes
the discussion of the solution quite a bit more cumbersome, as compared
to the case where hopping is absent.  Nevertheless, it is of course
important to gain insight whether the presence of such hopping
processes changes {\it all} the predictions of the idealized MCT or
whether there are only a few which are modified, since, after all, in
real materials such processes are always present.  We will see in the
next section to what extend the solutions of the mode-coupling
equations of the idealzed theory differ from the ones in which hopping
processes are taken into account.

\section*{Mode-Coupling Theory: Predictions and Tests}
\label{sec3}

In this section we will present some of the main predictions of MCT and
compare these with the results of experiments and computer simulations
of supercooled liquids. The results presented here are by no means
comprehensive in that the theory makes significantly more predictions,
quite a few of which have been tested in experiments or computer
simulations, which we do not discuss here. For a better overview the
reader is referred to the review articles~\cite{mct_rev} and the
collection of articles in Ref.~\cite{yip}.

\begin{figure}[hbt]
\epsfysize=7.0cm
\centerline{\epsfbox{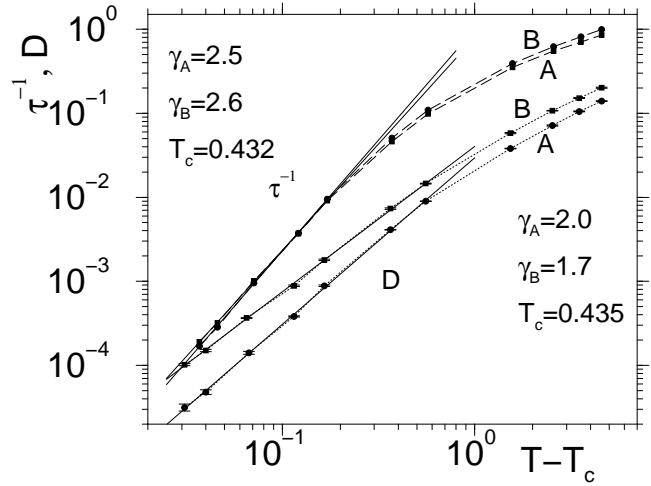}}
\caption{Temperature dependence of the diffusion constant $D$ and the
inverse relaxation times $\tau$ for the intermediate scattering
function for the two types of particles (type $A$ and type $B$) in a
binary Lennard-Jones system.  The straight lines are fits with a
power-law. From Ref.~[11]. 
}
\label{fig1}
\end{figure}

One of the main predictions of the idealized version of MCT is that
there exists a critical temperature $T_c$ at which the self diffusion
constant $D$, or the inverse of the $\alpha$-relaxation times $\tau$ of
the correlation functions, vanishes with a power-law, i.e.
\begin{equation}
D \propto \tau^{-1} \propto (T-T_c)^{\gamma}\quad,
\label{eq4}
\end{equation}
where the exponent $\gamma$ is larger than 1.5. Thus one can attempt to
see whether for a given system the low temperature dependence of these
quantities is given by such a law. In Fig.~\ref{fig1} we show that for
a binary Lennard-Jones system such a temperature dependence can indeed
be found and that the critical temperature $T_c$ is, in accordance with
MCT, independent of the quantity investigated. Furthermore MCT predicts that
the exponent $\gamma$ should be independent of the quantity
investigated. From the figure we recognize that this is reasonably well
fulfilled for this system if one compares the two relaxation times or
the two diffusion constants with each other, that, however, the
exponents for $D$ and $\tau^{-1}$ are definitely different. Thus this
prediction of the theory does not seem to hold for this particular
system.

\begin{figure}[hbt]
\epsfysize=7.0cm
\centerline{\epsfbox{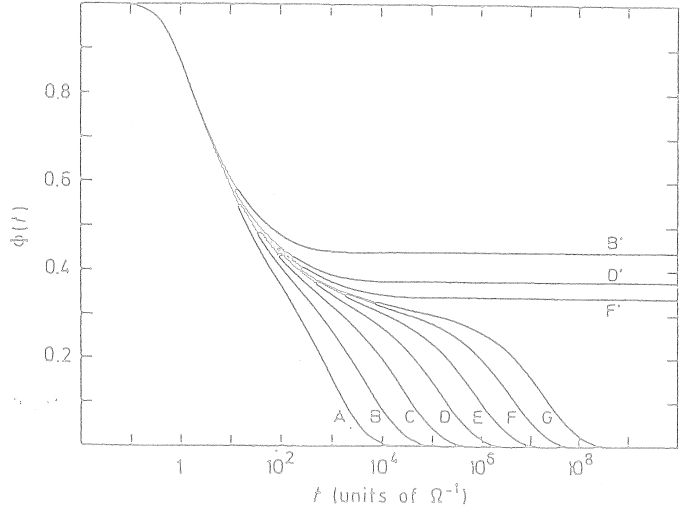}}
\caption{Time dependence of the correlation functions as computed from
a schematic model without hopping processes. The different curves
correspond to different values of the coupling parameters (see text for
details). From Ref.~[12] 
by permission.
}
\label{fig2}
\end{figure}

It is very instructive to study the full time and temperature
dependence of the solution of schematic models of the form given in
Eq.~(\ref{eq3}), since they can be compared with the relaxation
dynamics of real systems and thus allow to perform a more stringent
test of the theory. In Fig.~\ref{fig2} we show the correlation
functions for a model with $m(\Phi)=\lambda_1 \Phi+\lambda_2 \Phi^2$,
with $\lambda_i>0$, that were computed by G\"otze and
Sj\"ogren~\cite{gotze88} (with no hopping processes). The different
curves correspond to different values of the coupling parameters
$\lambda_i$ and are chosen such that their distance to the critical
values decreases like $0.2/2^n$ for $n=0,1,\ldots$ (liquid, curves
A,B,C \ldots) and increases like $0.2/2^n$ for $n=0,1,\ldots$ (glass,
curves F', D', \ldots).  Note that what in real systems corresponds to
a change in temperature corresponds in these schematic models to a
change in the coupling constants $\lambda_i$. In order to simplify the
language we will, however, in the following always use temperature as
the quantity that is changed.

From Fig.~\ref{fig2} we recognize that at short times the correlation
functions show a quadratic dependence on time, which is due to the
ballistic motion of the particles on these time scales. For high
temperatures, curve A, this relaxation behavior crosses directly over
to one which can be described well by an exponential decay. If the
temperature is decreased, curve D, there is an intermediate time
regime, the so called $\beta$-relaxation regime, where the correlation
function decays only very slowly, i.e. the $\Phi$ versus log(t) plot
exhibits an inflection point. For even lower temperatures, curve G, the
correlation functions show in this regime a plateau (log(t)
$\approx$ 5).  Only for even longer times the correlation function
enters the so-called $\alpha$-relaxation regime, the time window in
which it finally decays to zero.

The reason for the existence of the plateau is the following: At very
short times the motion of a particle is essentially ballistic. After a
microscopic time the particle starts to realize that it is trapped by
the cage formed by its nearest neighbors and thus the correlation
function does not decay any more.  Only for much longer times, towards
the end of the $\beta$-relaxation, this cage starts to break up and the
particle begins to explore a larger and larger volume of space. This
means that the correlation function enters the time scale of the
$\alpha$-relaxation and resumes its decay.

The closer the temperatures is to the critical temperature $T_c$, the
more this $\beta$-relaxation region stretches out in time and the time
scale of the $\alpha$-relaxation diverges with the power-law given by
Eq.~(\ref{eq4}), until at $T_c$ the correlation function does not decay
to zero any more. Upon a further lowering of the temperature the height
of the plateau increases and the time scale for which it can be
observed moves to {\it shorter} times.  MCT predicts that for temperatures
below $T_c$, this increase in the height of the plateau is proportional
to $\sqrt{T_c-T}$, which was indeed confirmed by, e.g., neutron scattering
experiments on a polymer glass former~\cite{frick90}.

Apart from the existence of a critical temperature $T_c$, one of the
most important predictions of the theory is the existence of three
different relaxation processes, which we have already seen in curve G
of Fig.~\ref{fig2}.  The first one is just the trivial relaxation on
the microscopic time scale. Since on this time scale the microscopic
details of the system are very important, hardly any general
predictions can be made for this time regime. This is different for the
second and third relaxation processes, the aforementioned $\beta$- and
$\alpha$-relaxation. For these time regimes MCT makes quite a few
predictions, some of which we will discuss now. Note that the
predictions, as stated below, are correct only in leading order in
$\sigma=(T_c-T)/T_c$. The corrections to this asymptotic behavior
have recently been worked out for the case of a hard sphere system and
it was found that they can be quite significant~\cite{franosch97}. Thus
for a {\it quantitative} comparison between experiments and MCT, these
corrections should be taken into account.

For the $\beta$-relaxation regime MCT predicts that its time scale
$t_{\sigma}$ diverges upon approach to the critical temperature $T_c$
as
\begin{equation}
t_{\sigma} \propto |T-T_c|^{1/2a}\quad,
\label{eq5}
\end{equation}
with $0<a<1/2$. Note that this type of singularity is predicted to
exist {\it above and below} $T_c$.  This divergent time scale can be
seen in Fig.~\ref{fig2}, in that the inflection point of the
correlation function in the $\beta$-relaxation regime moves to larger
times when the temperature is decreased to $T_c+0$ and that the time it
takes the correlator to reach the plateau diverges when the temperature
is increased to $T_c-0$.  Light scattering experiments have shown that
the divergence of the time scale of the $\beta$-relaxation, as given by
Eq.~(\ref{eq5}), can indeed be found in real
materials~\cite{li92,megen94}.

Furthermore the theory predicts that in the $\beta$-relaxation regime
the {\it factorization property} holds, by which the following is
meant.  If we consider a real system, as opposed to a schematic model,
we have correlation functions $F(q,t)$ which correspond to different
values of $q$, see Eq.~(\ref{eq2}). The factorization property says now
that in the $\beta$-relaxation regime these space-time correlation
functions can be written as
\begin{equation}
F(q,t)=f_c(q)+h(q) \sqrt{\sigma} g_{\pm}(t/t_{\sigma})\quad,
\label{eq5a}
\end{equation}
where the ($q$-dependent) constant $f_c(q)$ is the height of the
plateau, and is also called {\it nonergodicity parameter}, the
amplitude $h(q)$ is independend of temperature and time, and the $\pm$
in $g_{\pm}$ corresponds to $\sigma \stackrel{>}{<} 0$.  Thus the 
time dependence
of the correlation function enters only through the $q$-{\it
independent} function $g_{\pm}(t)$ which is therefore, for a given
system, ``universal''.  Equation~(\ref{eq5a}) means that, in the
$\beta$-relaxation regime, the time correlations are completely
independent from the spatial correlation, which is in stark contrast to
other types of dynamical processes such as, e.g., diffusion, where the
relaxation time of a mode for wave vector $q$ is proportional to
$q^{-2}$.

\begin{figure}[b]
\epsfysize=7.0cm
\centerline{\epsfbox{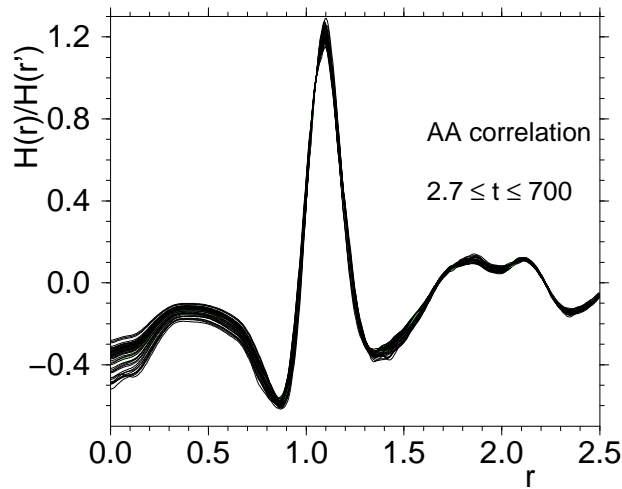}}
\caption{Ratio of critical amplitudes for different times in the
$\beta$-relaxation regime for the distinct part of the van Hove
correlation function in a binary Lennard-Jones system. From
Ref.~[17]. 
}
\label{fig3}
\end{figure}

In order to check whether for a given system the factorization
property holds, one can consider the space-Fourier transform of
Eq.~(\ref{eq5a}) which gives $F(r,t)=f_c(r)+H(r)\sqrt{\sigma}g_{\pm}(t)$. 
Making use of this last equation, it is simple to show that the ratio
\begin{equation}
\frac{F(r,t)-F(r,t')}{F(r',t)-F(r',t')}=\frac{H(r)}{H(r')}
\label{eq5b}
\end{equation}
is independent of time (here $r$ and $r'$ are arbitrary and $t$ and
$t'$ are times in the $\beta$-relaxation regime). In Fig.~\ref{fig3}
we show this ratio for the distinct part of the van Hove correlation
function of the binary Lennard-Jones system mentioned above. Every
curve corresponds to a different time $t$, all of which belong to the
$\beta$-relaxation regime. (The value of $t'$ is kept fixed at 3000
reduced time units and $r'=1.05$.) We see that all the curves lie in a
narrow band, which shows that the left hand side of Eq.~(\ref{eq5b}) is
indeed independent of time, i.e. that the factorization property holds.
Thus we see that for this system the time dependence of the correlation
functions are indeed given by a single function $g_{-}(t)$. The same
results were found for the dynamics of colloidal
suspensions~\cite{megen94}.

\begin{figure}[t]
\epsfysize=7.0cm
\centerline{\epsfbox{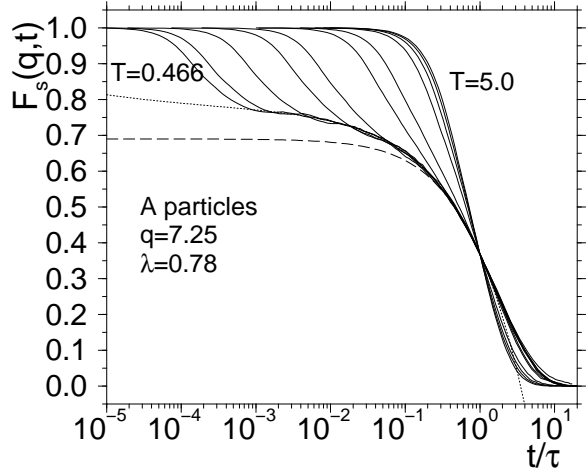}}
\caption{The incoherent intermediate scattering function $F_s(q,t)$
for a binary Lennard-Jones mixture versus rescaled time $t/\tau(T)$ for
different temperatures (solid lines). The dotted curve is a fit with
the functional form provided by MCT for the $\beta$-relaxation
regime.  The dashed curve is the result of a fit with the KWW function.
From Ref.~[18].
}
\label{fig4}
\end{figure}

It can be shown that the full time dependence of $g_{\pm}(t)$ can be
computed if one number $\lambda$, the so-called {\it exponent
parameter}, is known. This parameter can in turn be computed from the
structure factor, although such a computation is rather involved and
therefore $\lambda$ is often used as a fit parameter in order to fit
$g_{\pm}(t)$ to the data. The result of such a fit is shown in
Fig.~\ref{fig4} (dotted curve) where we show the incoherent
intermediate scattering function of the Lennard-Jones system discussed
above, versus $t/\tau(T)$. From this figure we recognize that for this
Lennard-Jones system the functional form provided by MCT is able to fit
the data very well in the late part of the $\beta$-relaxation regime.
However, for this system the {\it early} $\beta$-relaxation regime is
not fitted well by the $\beta$-correlator~\cite{kob95b}. The reason
for this is likely the strong influence of the microscopic dynamics.
This view is corroborated by the fact that if the dynamics is changed from a
Newtonian one to a stochastic one, the observed $\beta$-relaxation is
much more similar to the one predicted by the theory~\cite{gleim97}.
In addition to this, light scattering experiments have shown that in
colloidal systems the {\it whole} $\beta$-relaxation regime can be
fitted very well with the $\beta$-correlator~\cite{megen94,gotze91}
thus showing that there are systems for which the $\beta$-correlator
gives the correct description of the dynamics in the
$\beta$-relaxation regime.

The calculation of $g_{\pm}(t)$ from $\lambda$ is rather complicated and
therefore has to be done numerically. However, for fitting experimental
data, it is often more useful to have simple analytic expressions at
hand, even if they are correct only in leading order in $\sigma$, and MCT 
provides such expressions. It can be shown that in the {\it early}
$\beta$-relaxation regime, i.e. the time range during which the
correlator is already close to the plateau, but has not reached it yet,
the function $g_{\pm}(t)$ is a power-law, i.e.
\begin{equation}
g_{\pm}(t/t_{\sigma})=(t/t_{\sigma})^{-a} \quad , t/t_{\sigma} \ll 1 ,
\label{eq6}
\end{equation}
where the exponent $a$ is the same that appears in Eq.~(\ref{eq5}).
This time dependence is often also called {\it critical decay}.

For times in the {\it late} $\beta$-relaxation regime, i.e. in the time
interval where the correlator has already dropped below the plateau
but is still in its vicinity, MCT predicts that $g_{-}(t)$ is given by a
different power-law, the so-called von Schweidler law:
\begin{equation}
g_{-}(t/t_{\sigma})=-B(t/t_{\sigma})^b \quad ,t/t_{\sigma} \gg 1 ,
\label{eq7}
\end{equation}
where the exponent $b$ ($0<b\leq 1$) is related to $a$ via the nonlinear 
equation
\begin{equation}
\Gamma^2(1-a)/\Gamma(1-2a)=\Gamma^2(1+b)/\Gamma(1+2b)\quad,
\label{eq8}
\end{equation}
and $\Gamma(x)$ is the $\Gamma$-function. Furthermore it can be shown
that the left (or the right) hand side of this equation is equal to the
exponent parameter $\lambda$, which we have introduced before.  Thus if
one knows one element of the set $\{a,b,\lambda\}$, this relation and
Eq.~(\ref{eq8}) can be used to compute the two other elements of this
set. Light scattering experiments have shown that both power-laws,
Eqs.~(\ref{eq6}) and (\ref{eq7}), can be observed in real materials and
that Eq.~(\ref{eq8}) is indeed satisfied~\cite{li92,megen94}.

We now turn our attention to the relaxation of the correlation
functions on the time scale of the $\alpha$-relaxation. One of the
main results of MCT concerning this regime is the so-called
{\it time-temperature superposition principle} (TTSP), also this
valid to leading order in $\sigma$. This means that 
the correlators for different temperatures can be collapsed onto a
master curve $\Psi$ if they are plotted versus $t/\tau$, where $\tau$ is the
$\alpha$-relaxation time, i.e.
\begin{equation}
\Phi(t)=\Psi(t/\tau(T)) \quad .
\label{eq9}
\end{equation}
As an example for a system in which the TTSP works very well we show in
Fig.~\ref{fig4} the time dependence of $F_s(q,t)$, the incoherent
intermediate scattering function, of the Lennard-Jones system discussed
above, versus rescaled time $t/\tau(T)$. In accordance with MCT, the
curves for the different temperatures fall onto a master curve, if the
temperature is low enough. Furthermore the theory predicts that this
master curve can be fitted well with the so-called
Kohlrausch-Williams-Watts (KWW) function,
$\Phi(t)=A\exp(-(t/\tau(T)^{\beta})$. The result of such a fit is
included in the figure as well (dashed curve) and from it we 
recognize that this
prediction of the theory holds true also.

Contrary to the situation in the $\beta$-relaxation regime, where the
time dependence of the different correlation function was governed by the
single function $g_{\pm}(t)$, MCT predicts that in the $\alpha$-regime the
relaxation behavior of different correlation function is not universal.
This means that not only the amplitudes $A$ in the KWW function, but
also the exponent $\beta$ will depend on the correlator considered.
That such dependencies indeed exist has been demonstrated in
experiments and computer simulations~\cite{kob95b,mezei87}.

Although MCT predicts that the shape of the relaxation curves will
depend on the correlation function considered, the theory also predicts
that all of them share a common property, namely that all the corresponding
relaxation times diverge at $T_c$ with a power-law whose exponent
$\gamma$ is independent of the correlator, see Eq.~(\ref{eq4}). Also
this prediction was confirmed in experiments and in computer
simulations~\cite{megen94,kob95b}.

Furthermore the theory also predicts the existence of an interesting
connection between the exponents $a$ and $b$, which are important for
the $\beta$-relaxation [see Eqs.~(\ref{eq6}) and (\ref{eq7})], and the
exponent $\gamma$, which governs the time scale of the
$\alpha$-relaxation [see Eq.~(\ref{eq4})], in that
\begin{equation}
\gamma=1/2a+1/2b
\label{eq10}
\end{equation}
should hold. Thus, according to MCT, from measurements of the
temperature dependence of the $\alpha$-relaxation time we can learn
something about the time dependence of the relaxation in the
$\beta$-relaxation regime and vice versa.

\begin{figure}[b]
\epsfysize=7.0cm
\centerline{\epsfbox{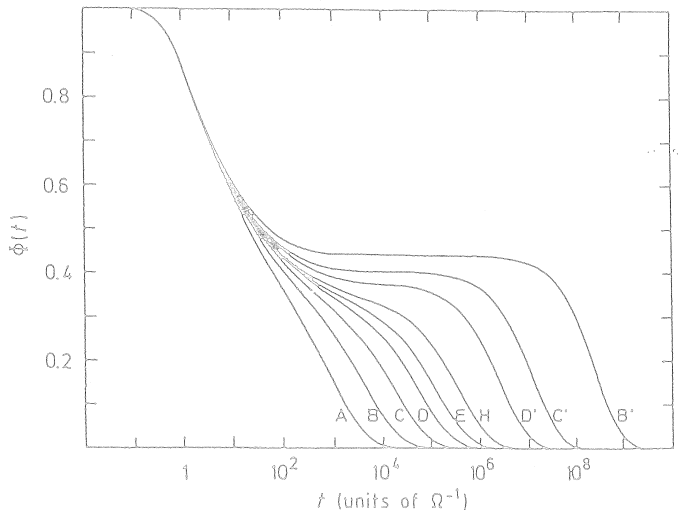}}
\caption{The incoherent intermediate scattering function $F_s(q,t)$
for a binary Lennard-Jones mixture versus rescaled time $t/\tau(T)$
for different temperatures (solid lines). The dotted curve is a fit with
the functional form provided by MCT for the $\beta$-relaxation
regime.  The dashed curve is the result of a fit with the KWW
function.  From Ref.~[18].
}
\label{fig5}
\end{figure}

Before we conclude this section we return to the extended version of
MCT, i.e. that form of the theory in which the hopping processes are
taken into account. In Fig.~\ref{fig5} we show the solution of the
same schematic model which was discussed in the context of
Fig.~\ref{fig2}, but this time with the inclusion of hopping
processes~\cite{gotze88}.
From this figure we recognize that the main
effect of such processes is that the correlation functions decay to
zero at {\it all} temperatures, which shows that the system is always
ergodic. Thus one {\it might} conclude that the concept of a critical
temperature does not make sense anymore, since there is no temperature
at which the relaxation times diverge. This is, however, not the case.
If the hopping processes are not too strong, there still will exist a
temperature range in which the relaxation times will show a power-law
behavior with a critical temperature $T_c$.  However, this power-law
will not extend down to $T_c$ but deviations will be observed in the
vicinity of $T_c$, in that the temperature dependence of $\tau$ will
be weaker than a power-law. Thus, despite the presence of the hopping
processes it is still possible to identify a $T_c$.  

\begin{figure}[b]
\epsfysize=7.0cm
\centerline{\epsfbox{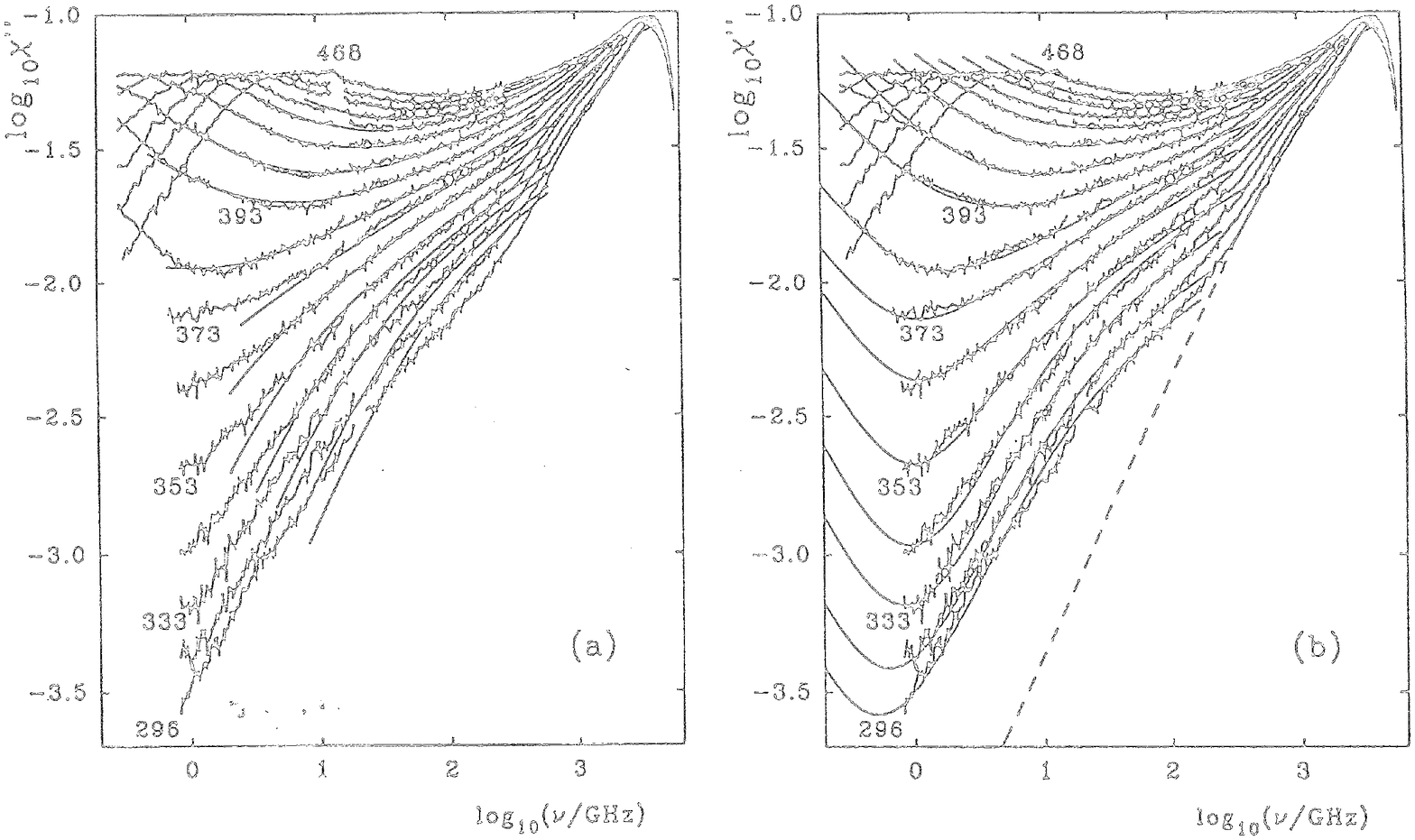}}
\caption{Susceptibility spectra of CKN and fits from MCT without
hopping processes (a) and with hopping processes (b). From
Ref.~[22] 
by permission.
}
\label{fig6}
\end{figure}

As a comparison between the corresponding curves in Figs.~\ref{fig2}
and \ref{fig5} shows, also the relaxation behavior of the correlation
functions are not affected too much by the hopping processes, {\it if
one is not too close to $T_c$}. Therefore many of the predictions that
MCT makes for the relaxation behavior hold even in the presence of such
processes. As an example of how important it can be to take into
account the hopping processes for the interpretation of real data very
close, or below, $T_c$, we show in Fig.~\ref{fig6} the results of a
light scattering experiment on calcium potassium nitrate (CKN) in a
temperature range which includes $T_c=378$ K~\cite{cummins93}. Shown is
the imaginary part of the susceptibility, i.e. the time Fourier
transform of the intermediate scattering function, multiplied by the
frequency $\omega$. In the left figure the data is analyzed by using
the idealized version of MCT and we recognize that although the theory
(smooth curves) fits the data (wiggly curves) well for temperature
above $T_c$, significant deviations occur at lower temperatures, since
there the hopping processes become important. If these are taken into
account in a phenomenological way, the agreement between theory and
experiment is very good for the whole temperature range (right part of
the figure). It has to be noted that in the latter set of fits the
theory makes use of one additional fit parameter, the strength of the
hopping processes. Nevertheless, the inclusion of the hopping processes
has improved the agreement between theory and experiment in such a
dramatic way, that the price of one additional fit parameter seems to
be well justified.

\section*{Conclusions}
\label{sec4}

The goal of this article was to give a concise introduction to the
mode-coupling theory of the glass transition. 
For this we briefly described the origin of the
theory and explained some of its main predictions. Although we have
presented here only a few results of the tests by which the capability
of the theory to describe the dynamics of supercooled liquid were
investigated, we mention that in the last few years many other such
tests were performed~\cite{yip}. Most of them showed that the 
theory is {\it at
least} able to describe {\it certain} aspects of this dynamics and that
there exist systems for which the predictions of MCT are correct in
surprisingly great detail. One unexpected result that came out of these tests
is that MCT seems to work reasonably well even for systems that are
very different (e.g. polymers) from the simple liquids for which the
theory was originally devised. Thus one might tentatively conclude that
the basic mechanism that leads to the slowing down of a liquid upon
supercooling is not specific to simple liquids and can be described
with the help of MCT.

It is interesting to note that most of the systems for which MCT gives
a good description of the dynamics belong to the class of fragile glass
formers~\cite{angell85}, i.e. are systems that show a significant
change of their activation energy if their viscosity is plotted in an
Arrhenius plot. This bend occurs at a temperature that is about 10-40 K
above the glass transition temperature and investigations of the
dynamics of these systems have shown that the $T_c$ of MCT is in the
vicinity of this bend. In the ideal version of MCT, $T_c$ corresponds
to the temperature at which the system undergoes a structural arrest.
Since in experiments it is found that at the temperature $T_c$ the
viscosity of the systems is significantly enhanced with respect to the
one at the triple point but by no means large, we thus must conclude
that for most real systems the hopping processes, which are neglected
in the idealized theory, become important in the vicinity of $T_c$.
Therefore for temperatures in the vicinity of $T_c$ the extended
version of the theory has to be used. However, despite the presence of
these hopping processes, the signature of the sharp transition of the
idealized theory is still observed and gives rise to a dynamical
anomaly in the relaxation behavior of the system, e.g. the pronounced
bend in the viscosity. Thus from the point of view of MCT the dynamics
of supercooled liquids can be described as follows: In the temperature
range where the liquid is already supercooled, but which is still above
$T_c$, the dynamics is described well by the idealized version of MCT.
In the vicinity of $T_c$ the extended version of the theory has to be
used, whereas for temperatures much below $T_c$ it is likely that the
simple way MCT takes into account the hopping processes is no longer
adequate and a more sophisticated theory has to be used.

From the above one might get the impression that MCT is applicable
only for fragile glass formers and not for strong ones. That this is
not necessarily the case is shown by a recent calculations by Franosch
{\it et al.} who have shown that also features in the dynamics of
glycerol, a glass former that is rather strong, can be described well
by simple schematic models~\cite{franosch96b}. Whether this is also
true for very strong glass formers such as SiO$_2$ remains to be
seen, however.

In this article we have shown that MCT is able to give a {\it
qualitative} correct description of the dynamics of certain supercooled
liquids in that, e.g., the relaxation of system in the $\beta$-regime
shows the critical decay or the von Schweidler law predicted by the
theory. To this two important comments have to be added: The first one
is that the predictions of the theory that have been presented here are
valid only asymptotically close to $T_c$. If one considers temperatures
that have a finite distance from $T_c$, corrections to the mentioned
scaling laws become important. Some of these corrections have very
recently been computed for a model of hard spheres and it was found
that they can be quite important in order to make a {\it consistent}
analysis of experimental data within the framework of
MCT~\cite{franosch97}. The second comment we make is, that MCT is not
only able to make {\it qualitative} statements on the relaxation
behavior of supercooled liquids but that it is also able to predict the
non-universal values of the parameters of the theory, such as $T_c$,
the exponent parameter or the $q$-dependence of the nonergodicity
parameter, reasonably well~\cite{megen94,gotze91,nauroth97}.  Unfortunately
these last types of calculations are rather involved, since one has to
take into account the full wave-vector dependence of the mode-coupling
equations, and thus have been done only for a few cases.  Despite these
difficulties it would be very useful to have more investigations of
this kind since they allow to test the range of applicability of the
theory to a much larger extend that the tests of the ``universal''
predictions of the theory allow.

Acknowledgments: We thank W. G\"otze for many useful comments on this
manuscript. Part of this work was supported by the DFG under SFB
262/D1.

\end{document}